\documentclass[a4paper, 11pt]{article}
\usepackage{tikz} % Package for drawing
\usepackage{amsmath}
\usepackage{amssymb}
\usepackage{hyperref}
\usepackage{listings}
\usepackage{dsfont}
\usepackage{graphicx}
\usepackage{subfigure}
\usepackage{multirow}
\usepackage[round]{natbib} % enables pretty citations
\usepackage{enumitem}% enables pretty lists
\usepackage{authblk}
\usepackage[margin=1in]{geometry}
\usepackage{algorithm,algpseudocode}
\usepackage{bm}
\usepackage{tabu} 
\usepackage{multirow}
\usepackage{comment}
\usepackage{xspace}
\usepackage{nicefrac}
\usepackage{lipsum}
\usepackage{booktabs}
\usepackage{lineno}
\usetikzlibrary{tikzmark}
\newcommand{\E}{\mathbb{E}}
\newcommand{\SparseSignatures}{\texttt{SparseSignatures}\xspace}
\newcommand{\SignatureAnalyzer}{\texttt{SignatureAnalyzer}\xspace}
\newcommand{\SigneR}{\texttt{SigneR}\xspace}

\newcommand\pef[1]{(\ref{#1})}

\newcommand*\justify{%
  \fontdimen2\font=0.4em% interword space
  \fontdimen3\font=0.2em% interword stretch
  \fontdimen4\font=0.1em% interword shrink
  \fontdimen7\font=0.1em% extra space
  \hyphenchar\font=`\-% allowing hyphenation
}

\renewcommand{\texttt}[1]{%
  \begingroup
  \ttfamily
  \begingroup\lccode`~=`/\lowercase{\endgroup\def~}{/\discretionary{}{}{}}%
  \begingroup\lccode`~=`[\lowercase{\endgroup\def~}{[\discretionary{}{}{}}%
  \begingroup\lccode`~=`.\lowercase{\endgroup\def~}{.\discretionary{}{}{}}%
  \catcode`/=\active\catcode`[=\active\catcode`.=\active
  \justify\scantokens{#1\noexpand}%
  \endgroup
}

\usepackage{tikz}
\usepackage{collcell}
\newcommand*{\MinNumber}{0}%
\newcommand*{\MaxNumber}{1}%
\definecolor{blue}{RGB}{64, 131, 252}
\definecolor{purple}{RGB}{154, 76, 149}
\definecolor{green}{RGB}{125, 206, 130}
\newcommand{\ApplyGradient}[1]{%
        \pgfmathsetmacro{\PercentColor}{100.0*(exp(#1)-exp(\MinNumber))/(exp(\MaxNumber)-exp(\MinNumber))}
        \hspace{-0.33em}\colorbox{green!\PercentColor!white}{\makebox(4,4){\small #1}}
}
 
\newcolumntype{R}{>{\collectcell\ApplyGradient}c<{\endcollectcell}}

\title{Model selection and robust inference of mutational signatures using Negative Binomial non-negative matrix factorization} 

\author{Marta Pelizzola$^1$, Ragnhild Laursen$^1$  and Asger Hobolth$^1$}
\date{  %
    {\small $^1$Department of Mathematics, Aarhus University. Emails: \{marta, ragnhild, asger\}@math.au.dk}\\%
    [2ex]%
    \today
}

\begin{document}
%%%%%%%%%%%%%%%%%%%%%%%%%%%%%%%%%%%%%%%%%%%
%%% Generate title
%%%%%%%%%%%%%%%%%%%%%%%%%%%%%%%%%%%%%%%%%%%
\maketitle
\vspace*{0.35in}

\section*{Abstract}

The spectrum of mutations in a collection of cancer genomes can be described by a mixture of a few mutational signatures. The mutational signatures can be found using non-negative matrix factorization (NMF). To extract the mutational signatures we have to assume a distribution for the observed mutational counts and a number of mutational signatures. In most applications, the mutational counts are assumed to be Poisson distributed, and the rank is chosen by comparing the fit of several models with the same underlying distribution and different values for the rank using classical model selection procedures. However, the counts are often overdispersed, and thus the Negative Binomial distribution is more appropriate. We propose a Negative Binomial NMF with a patient specific dispersion parameter to capture the variation across patients. We also introduce a novel model selection procedure inspired by cross-validation to determine the number of signatures. Using simulations, we study the influence of the distributional assumption on our method together with other classical model selection procedures and we show that our model selection procedure is more robust at determining the correct number of signatures under model misspecification. We also show that our model selection procedure is more accurate than state-of-the-art methods for finding the true number of signatures. Other methods are highly overestimating the number of signatures when overdispersion is present. We apply our proposed analysis on a wide range of simulated data and on two real data sets from breast  and prostate cancer patients. The code for our model selection procedure and negative binomial NMF is available in the R package SigMoS and can be found at \url{https://github.com/MartaPelizzola/SigMoS}.

\textbf{Keywords:} cancer genomics, cross-validation, model checking, model selection, mutational signatures, Negative Binomial, non-negative matrix factorization, Poisson.

\textbf{AMS classification: 92-08}, 92-10, 62-08 

%%%%%%%%%%%%%%%%%%%%%%%%%%%%%%%%%%%%%%%%%%%
%%% Introduction
%%%%%%%%%%%%%%%%%%%%%%%%%%%%%%%%%%%%%%%%%%%
\section{Introduction}
Somatic mutations occur relatively often in the human genome and are mostly neutral. However, the accumulation of some mutations in a genome can lead to cancer. The summary of somatic mutations observed in a tumor is called a mutational profile and can often be associated with factors such as aging \citep{Risques2018}, UV light \citep{Shibai2017} or tobacco smoking \citep{Alexandrov2016}. A mutational profile is thus a mixture of mutational processes that are represented by mutational signatures. Several signatures have been identified from the mutational profiles and associated with different cancer types \citep{Alexandrov2020,Tate2019}.

A common strategy to derive the mutational signatures is non-negative matrix factorization \citep{alexandrov2013breastcancer, Nik-Zainal2012, Lal2021}. Non-negative matrix factorization (NMF) is a factorization of a given data matrix $V \in \mathbb{N}_0^{N \times M}$ into the product of two non-negative matrices $W \in \mathbb{R}_+^{N \times K}$ and $H \in \mathbb{R}_+^{K \times M}$ such that
\[V \approx WH.\]
The rank $K$ of the lower-dimensional matrices $W$ and $H$ is much smaller than $N$ and $M$.

In cancer genomics, the data matrix $V$ contains the mutational counts for different patients, also referred to as mutational profiles. The number of rows $N$ is the number of patients and the number of columns $M$ is the number of different mutation types. In this paper, $M = 96$ corresponding to the 6 base mutations, when assuming strand symmetry times the 4 flanking nucleotides on each side ( i.e. $4 \cdot 6 \cdot 4 = 96$). The matrix $H$ consists of $K$ mutational signatures defined by probability vectors over the different mutation types. In the matrix $W$, each row contains the weights of the signatures for the corresponding patient. In this context, the weights are usually referred to as the exposures of the different signatures.

To estimate $W$ and $H$ we need to choose a model and a rank $K$ for the data $V$. These two decisions are highly related as the optimal rank of the data $V$ is often chosen by comparing the fit under a certain model for many different values of $K$. The optimal $K$ is then found using a model selection procedure such as Akaike Information Criterion (AIC), Bayesian Information Criterion (BIC) or similar approaches described in Section \ref{sec:methods_CV}. Most methods used in the literature \citep{alexandrov2013breastcancer,Fischer2013,Rosales2017} for choosing the rank are based on the likelihood value, which depends on the assumed model. 
For mutational counts the usual model assumption is the Poisson distribution \citep{alexandrov2013breastcancer}
\begin{align} \label{eq:poisson}
    V_{nm} \sim \text{Po}((WH)_{nm}),
\end{align}
where $W$ and $H$ are estimated using the algorithm from \cite{Lee1999} that minimizes the generalized Kullback-Leibler divergence. The algorithm is equivalent to maximum likelihood estimation, as the negative log-likelihood function for the Poisson model is equal to the generalized Kullback-Leibler up to an additive constant. We observe that this model assumption often leads to overdispersion for mutational counts, i.e. to a situation where the variance in the data is greater than what is expected under the assumed model.

We therefore suggest using a model where the mutational counts follow a Negative Binomial distribution that has an additional parameter to explain the overdispersion in the data. The Negative Binomial NMF is discussed in \cite{Gouvert2020}, where it is applied to recommender systems, and it has recently been used in the context of cancer mutations in \cite{Lyu2020}. They apply a supervised Negative Binomial NMF model to mutational counts from different cancers which uses cancer types as metadata. Their aim is to obtain signatures with a clear etiology, which could be used to classify different cancer types. 

For mutational count data we extend the Negative Binomial NMF model by including patient specific dispersion. The extended model is referred to as NB$_\text{N}$-NMF, where $N$ is the number of dispersion parameters. We investigate when and why NB$_\text{N}$-NMF is more suitable for mutational counts than the usual Poisson NMF (Po-NMF). In particular we evaluate the goodness of fit for mutational counts using a residual-based approach. Despite the above mentioned recent efforts, we still believe, as it has also been mentioned in \cite{Fevotte2009}, that a great amount of research has been focusing on improving the performance of NMF algorithms given an underlying model and less attention has been directed to the choice of the underlying model given the data and application.

Since the number of signatures depends on the chosen distributional assumption, we suggest using NB$_\text{N}$-NMF and we propose a novel model selection framework to choose the number of signatures. We show that our model selection procedure is more robust toward
inappropriate model assumptions compared to other methods currently used in the literature such as \SigneR, \SignatureAnalyzer and \SparseSignatures. We use both simulated and real data to validate our proposed model selection procedure against other methods.

We have implemented our methods in the R package SigMoS that includes NB$_N$-NMF and the model selection procedure. The R package is available at \url{https://github.com/MartaPelizzola/SigMoS}. The package also contains the simulated and real data used in this paper.

\section{Negative Binomial non-negative matrix factorization}\label{sec:NegBinNMF}
In this section we first argue why the Negative Binomial model in \cite{Gouvert2020} is a natural model for the number of somatic mutations in a cancer patient. Then we describe our patient specific Negative Binomial non-negative matrix factorization NB$_\text{N}$-NMF model and the corresponding estimation procedure. 

\subsection{Negative Binomial model for mutational counts}\label{subsec:negbin}
We start by illustrating the equivalence of the Negative Binomial to the more natural Beta-Binomial model as a motivation for our model choice. Assume a certain mutation type can occur in $\tau$ triplets along the genome with a probability $p$. Then it is natural to model the mutational counts with a binomial distribution \citep{Weinhold2014, Lochovsky2015}
\begin{equation} \label{eq:betadist}
    V_{nm} \sim \text{Bin}(\tau,p).
\end{equation}
However, \cite{Lawrence2013} observed that the probability of a mutation varies along the genome and is correlated with both expression levels and DNA replication timing. We therefore introduce the Beta-Binomial model  
\begin{align} \label{eq:betabinom}
\begin{split}
   V_{nm}|&p \sim \text{Bin}( \tau, p) \\ 
    &p \sim \text{Beta}(\alpha, \beta),
    \end{split}
\end{align}
where the beta prior on $p$ models the heterogeneity of the probability of a mutation for the different mutation types due to the high variance along the genome. As $p$ follows a Beta distribution, its expected value is $\E[p] = \nicefrac{\alpha}{(\alpha + \beta)}$. 
For mutational counts, the number of triplets $\tau$ is extremely large and the probability of mutation $p$ is very small. In the data described in \cite{Lawrence2013} there are typically between 1 and 10 mutations per megabase with an average of 4 mutations per megabase ($\tau \approx 10^6$). This means $\E[p] = \nicefrac{\alpha}{(\alpha + \beta)} \approx 4 \cdot 10^{-6}$ and thus, for mutational counts in cancer genomes we have that $\beta >> \alpha$. 
As $\tau$ is large and $p$ is small, the Binomial model is very well approximated by the Poisson model
$\text{Bin}(\tau, p) \backsimeq \text{Pois}(\tau p)$.
This distributional equivalence of Poisson and Binomial when $\tau$ is large and $p$ is small is well known. This also means that the models \eqref{eq:poisson} and \eqref{eq:betadist} are approximately equivalent with $\tau p = (WH)_{nm}$. 

The Beta and Gamma distributions are also approximately equivalent in our setting. Indeed, as $\beta >> \alpha$, the Beta density can be approximated by the Gamma density in the following way 
\begin{align*}
\frac{p^{\alpha - 1}(1-p)^{\beta - 1}}{B(\alpha, \beta)} = \frac{p^{\alpha - 1}}{\Gamma(\alpha)} (\beta-1+\alpha)(\beta-1+(\alpha-1)) \cdots (\beta-1) (1-p)^{\beta - 1} \approx \frac{p^{\alpha - 1}}{\Gamma(\alpha)} \beta^\alpha (e^{-p})^{\beta}.
\end{align*}
Therefore, for mutational counts the model in \eqref{eq:betabinom} is equivalent to
\begin{align} \label{eq:gammapoisson}
\begin{split}
   V_{nm}|&p \sim \text{Pois}( \tau p) \\ 
    &p \sim \text{Gamma}(\alpha, \beta).
\end{split}
\end{align}
Since the Negative Binomial model is a Gamma-Poisson model we can also write the model as
\[ V_{nm} \sim \text{NB} \left(\alpha, \frac{\tau }{\beta + \tau}\right) \backsimeq \text{NB} \left(\alpha, \frac{ \tau \E[ p]}{\alpha + \tau \E[p]}\right) \backsimeq \text{NB}\left(\alpha, \frac{(WH)_{nm}}{\alpha + (WH)_{nm}} \right),\]
where the last parametrization is equivalent to the one in \cite{Gouvert2020}. In the first distributional equivalence we use $\E[p] \approx \frac{\alpha}{\beta}$ and in the second we use $\tau \E[ p] = (WH)_{nm}$. Compared to the Beta-Binomial model, the Negative Binomial model has one fewer parameter and is analytically more tractable. The mean and variance of this model are given by 
\begin{align} \label{eq:mean_var_nb}
\E[V_{nm}] = (WH)_{nm} \quad \text{and} \quad \text{Var}(V_{nm}) = (WH)_{nm} \left(1 + \frac{(WH)_{nm}}{\alpha}\right). 
\end{align}
In recent years, this model has become more popular to model the dispersion in mutational counts \citep{Martincorena2017,Zhang2020}. When $\alpha \rightarrow \infty$ above, the Negative Binomial model converges to the more commonly used Poisson model as $\text{Var}(V_{nm}) \downarrow (WH)_{nm}$. As shown in this section, the Negative Binomial model can be seen both as an extension of the Poisson model and as equivalent to the Beta-Binomial model.  Thus, we opted to implement a  negative binomial NMF model for mutational count data. More details on the approximation of the Negative Binomial to the Beta-Binomial distribution can also be found in \cite{Teerapabolarn2015}.

\subsection{Patient specific Negative Binomial NMF: NB$_\text{N}$-NMF} \label{subsec:patientNBNMF}
\cite{Gouvert2020} and \cite{Lyu2020} present a Negative Binomial model where $\alpha$ is shared across all observations. However, the probability of a mutation in \eqref{eq:betabinom} is also varying across patients (see e.g. mutational burden in \cite{RareSignatures2022}), thus we extend the model by allowing patient specific dispersion. We assume 
\[V_{nm} \sim \text{NB}\left( \alpha_n, \frac{(WH)_{nm}}{\alpha_n + (WH)_{nm}}\right) \] where $n \in 1, \dots, N$ correspond to the different patients. As for the estimation of $W$ and $H$ in \cite{Gouvert2020}, we define the following divergence measure:
\begin{align}
    d_N(V||WH) & = \sum_{n=1}^N \sum_{m=1}^M \left \{ V_{nm} \log \left(\frac{V_{nm}}{ (WH)_{nm}}\right) - (\alpha_n + V_{nm}) \log \left(\frac{\alpha_n + V_{nm}}{\alpha_n + (WH)_{nm}} \right) \right \}. \label{eq:div}
\end{align} 
In Section \ref{sec:methods} we show that the negative of the log-likelihood function is equal to this divergence up to an additive constant.
Indeed, this is a divergence measure as $d_N(V||WH) = 0$ when $V = WH$ and $d_N(V||WH)>0$ for $V \neq WH$. We can show this by defining $g(t) = (V_{nm}+t)\log \left( \nicefrac{ (V_{nm}+t)}{((WH)_{nm}+t) }\right)$ and realize $d_N(V||WH) = g(0) - g(\alpha) \geq 0$ because $g'(t)\leq 0$ with equality only when $V = WH$. % and $\alpha > 0$.
In our application, we find maximum likelihood estimates (MLEs) of $\alpha_1, \dots , \alpha_N$ based on the Negative Binomial likelihood using Newton-Raphson together with the estimate of $WH$ from Po-NMF. We opted for this more precise estimation procedure for $\alpha_1, \dots , \alpha_N$ instead of the grid search approach used in \cite{Gouvert2020}. Final estimates of $W$ and $H$ are then found by minimizing the divergence in \eqref{eq:div} by the iterative majorize-minimization procedure (see the derivation in Section \ref{sec:methods}).
The NB$_\text{N}$-NMF procedure is described in Algorithm \ref{alg:nbnmf_alphan} and further details can be found in Section \ref{sec:methods_alphai}. The model in \cite{Gouvert2020} and \cite{Lyu2020} is similar except $\alpha_1 = \cdots = \alpha_N = \alpha$.
\begin{algorithm}[h!]
\caption{NB$_\text{N}$-NMF: Estimation of $W$, $H$ and $\{ \alpha_1, \dots, \alpha_N \}$}\label{alg:nbnmf_alphan}
\begin{algorithmic}[1]
\Require{$V, K, \epsilon$}
\Ensure{$W$, $H$, $\{ \alpha_1, \dots, \alpha_N \}$ } 
\State $W^{Po}, H^{Po} \gets$ apply Po-NMF to $V$ with $K$ signatures
\State $\alpha_1, \dots , \alpha_N \gets$ Negative Binomial MLE using $W^{Po}, H^{Po}$ and $V$
\State Initialize $W^1,H^1$ from a random uniform distribution
\For{$i = 1, 2, \dots$} 
\State $W^{i+1}_{nk} \gets W^i_{nk} \frac{\sum_{m=1}^M \frac{V_{nm}}{(W^iH^i)_{nm}} H^i_{km}}{\sum_{m=1}^M \frac{V_{nm} + \alpha_n}{(W^iH^i)_{nm} + \alpha_n} H^i_{km}}$
\linespread{2.7}\selectfont
\State $H^{i+1}_{km} \gets H_{km}^i \frac{\sum_{n=1}^N \frac{V_{nm}}{(W^{i+1}H^i)_{nm}} W^{i+1}_{nk}}{\sum_{n=1}^N \frac{V_{nm} + \alpha_n}{(W^{i+1}H^i)_{nm} + \alpha_n} W^{i+1}_{nk}}$
\linespread{2.7}\selectfont
\If{$|d_N(V||W^{i+1}H^{i+1}) - d_N(V||W^{i}H^{i}) | < \epsilon$}
\State \Return $ W, H \gets W^{i+1}, H^{i+1}$
\linespread{1}\selectfont
\EndIf
\EndFor
\end{algorithmic}
\end{algorithm}

\section{Estimating the number of signatures} \label{sec:methods_CV}
Estimating the number of signatures is a difficult problem when using NMF. More generally, estimating the number of components for mixture models or the number of clusters is a well known challenge in applied statistics. 

Examples of the complexity of this problem can be found in the $K$-means clustering algorithm and in Gaussian mixture models where the number of clusters $K$ has to be provided for the methods. The silhouette and the elbow method are among the most common techniques to estimate $K$ for $K$-means clustering, however it is often unclear how to find an exact estimate of $K$. A detailed description of these challenges can be found in \cite{Gupta2018}. Here the authors also propose a new way of estimating the number of clusters that follows the same rationale as the elbow method, but it combines the detection of optimal well-separated clusters and clusters with equal number of elements. The discrepancy between these two solutions is then used to determine $K$. 

Estimating the number of components is also a critical issue for mixed membership models. One example can be found in the estimation of the number of subpopulations in population genetics. Population structure is indeed modeled as a mixture model of $K$ subpopulations and the inference of $K$ is challenging. In \cite{Pritchard2000} an ad hoc solution is proposed under the assumption that the posterior distribution follows a normal distribution, which is often violated in practice. \cite{Verity2016} take a different approach and derive a new estimator using thermodynamic integration based on the "power posterior" distribution. This is nothing more than the ordinary posterior distribution, but with the likelihood raised to a power to ensure that the distribution integrates to 1. This procedure seems to be very accurate, however it is computationally intense and thus can only be used on small data sets.

Classical procedures to perform model selection are the Akaike Information Criterion (AIC)
\begin{align}\label{eq:AIC}
     \text{AIC} = -2\ln L + 2n_{prm}
\end{align}
and the Bayesian Information Criterion (BIC)
\begin{align}\label{eq:BIC}
     \text{BIC} = -2\ln L + \ln(n_{obs}) n_{prm}
\end{align}
where $\ln L$ is the estimated log-likelihood value, $n_{obs}$ is the number of observations and $n_{prm}$ the number of parameters to be estimated. The two criteria attempt to balance the fit to the data (measured by $-2\ln L$) and the complexity of the model (measured by the scaled number of free parameters). We have $n_{obs} = N$ where $N$ is the number of patients, so $\ln(n_{obs}) > 2$ if $N \geq 8$, which means that in our context the number of parameters has a higher influence for BIC compared to AIC because real data sets always have at least tens of patients. Additionally, the structure of the data matrix $V$ can lead to two different strategies for choosing $n_{obs}$ when BIC is used. Indeed, the number of observations in this context can be set as the total number of counts (i.e.\, $N \cdot M$) or as the number of patients $N$, leading to an ambiguity in the definition of this criterion. \cite{Verity2016} also presents results on the performance of AIC and BIC, where the power is especially low for BIC. AIC provides higher stability in the scenario from \cite{Verity2016}, however it does not seem suitable in our situation due to a small penalty term.

A very popular model selection procedure is cross-validation. In \cite{Gelman2013} they compare various model selection methods including AIC and cross-validation. Here, the authors recommend to use cross-validation as they demonstrate that the other methods fail in some circumstances. In \cite{Luo2017} they also show that cross-validation has better performance than the other considered methods, including AIC and BIC. 

\subsection{Model selection for NMF}

For NMF we propose an approach for estimating the rank which is highly inspired by cross-validation. As for classical cross-validation we split the patients in $V$ in a training and a test set multiple times.  

Since all the parameters in the model i.e. $W$ and $H$ are free parameters it means that the exposures for the patients in the test set are unknown from the estimation of the training set. The patients in the training set give an estimation of the signatures and the exposures of the patients in the training set. One could argue to fix the signatures from the training set and re-estimate exposures for the test set, but we observed that this lead to an overestimation of the test set.

Instead we have chosen to fix the exposures to the ones estimated from the full data. This means our evaluation on the test set is a combination of estimated signatures from the training set and exposures from the full data. The idea is to exploit the fact that the signature matrix should be robust to changes in the patients included in the training set. If the estimated signatures are truly explaining the main patterns in the data, then we expect the signatures obtained from the training set to be similar to the ones from the full data. Therefore the product of the exposures from the full data and the signatures from the training set should give a good approximation of the test set, if the number of signatures is appropriate.

Inputs for the procedure are the data $V$, an NMF method, the number of signatures $K$, the number of splits into training and test $J$ and the $cost$ function. We evaluate the model for a range of values of $K$ and then select the model with the lowest cost. The NMF methods we are using here are either Po-NMF from \cite{Lee1999} or NB$_N$-NMF in Algorithm \ref{alg:nbnmf_alphan}, but any NMF method could be applied.

\begin{figure}[H]
    \centering
    \includegraphics[width =0.7\textwidth]{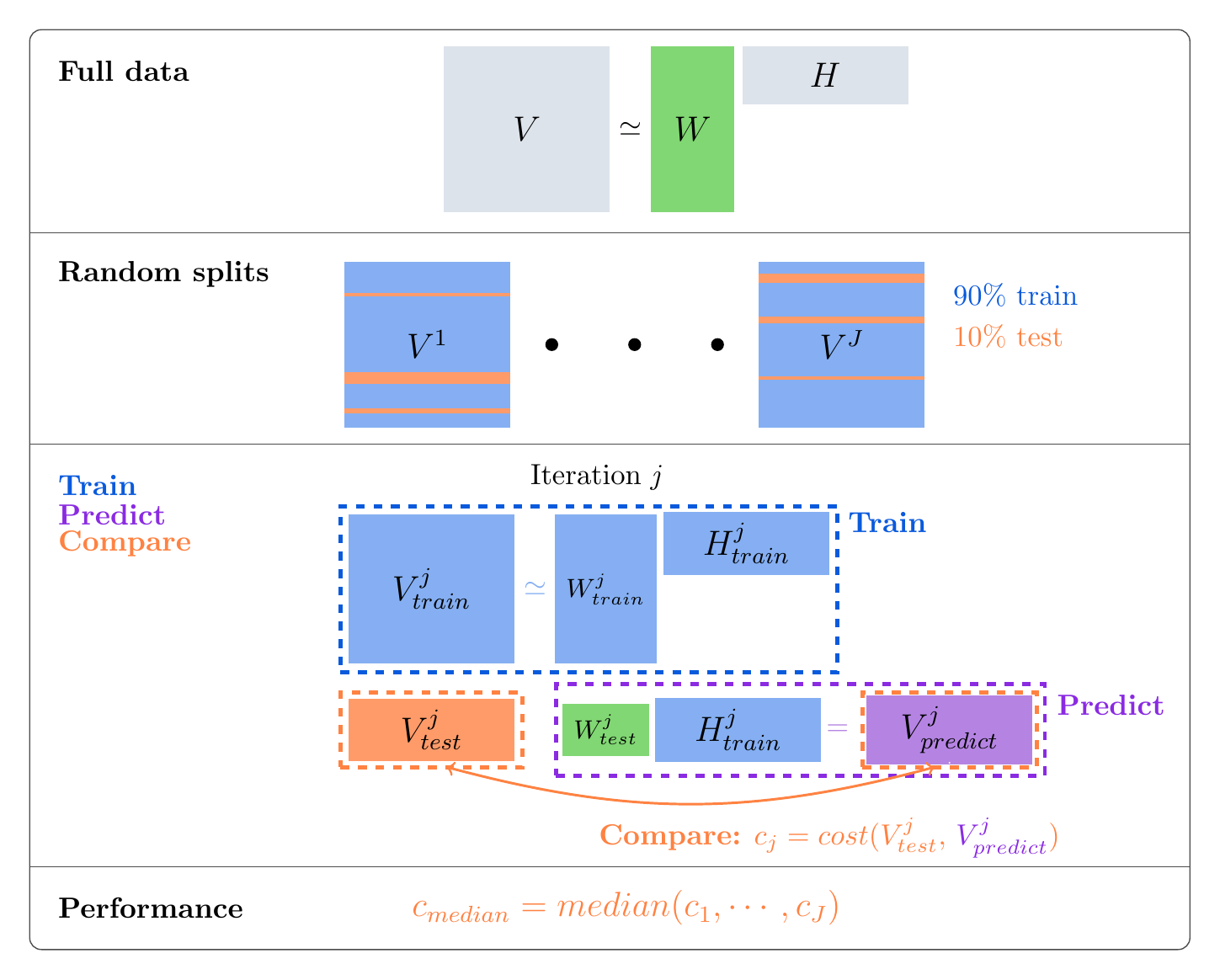}
    \vspace{-1em}
    \caption{SigMoS procedure for a given number of signatures $K$ and a count matrix $V$. Pseudocode can be found in Algorithm \ref{alg:crossval}.}
    \label{fig:cv_algorithm}
\end{figure}

A visualization of our model selection algorithm can be found in Figure \ref{fig:cv_algorithm}. First, we consider the full data matrix $V$ and we apply the chosen NMF algorithm to obtain an estimate for both $W$ and $H$. Afterwards, for each iteration, we sample $90\%$ of the patients randomly to create the training set and determine the remaining $10\%$ as our test set. We then apply the chosen NMF method to the mutational counts of the training set obtaining an estimate $W_{train}$ and $H_{train}$. 

Now, as for classical cross-validation, we want to evaluate our model on the test set. To evaluate the model here, we use the full data: indeed, we multiply the exposures relative to the patients in the test set  estimated on the full data $W_{test_j}$ times the corresponding signatures estimated from the training set $H_{train_j}$. We use the prediction of the test data to evaluate the model computing the distance between the true data $V_{test_j}$ and their prediction with a suitable $cost$ function. This procedure is iterated $J$ times leading to $J$ cost values $c_j$, $j=1, \dots, J$. The median of these values is calculated for each number of signatures $K$. We call this procedure SigMoS and summarize it in Algorithm~\ref{alg:crossval}. The optimal $K$ is the one with the lowest cost. We use the generalized Kullback-Leibler divergence as a cost function and discuss the choice of cost function in Section~\ref{sec:discussion}. We compare the influence of the model choice for our procedure to AIC and BIC. We also compare to \SignatureAnalyzer, \SigneR and \SparseSignatures as these are recently introduced methods in the literature and examine the results from this comparison in Section \ref{sec:results_simulations}.

\mbox{}

\begin{algorithm}[h]
\caption{SigMoS: Cost for a given number of signatures $K$ for the count matrix $V$}\label{alg:crossval}
\begin{algorithmic}[1]
\Require{$V, K, J, cost,$ NMF-method}
\Ensure{$c_{median}$} 
\State $W, H \gets$ apply the chosen NMF method to $V$ with $K$ signatures
\For{$j = 1$ to $J$} 
\State $V_{train}^j \gets $ mutational counts for the patients in the $j^{th}$ training set
\State $V_{test}^j \gets V \setminus V_{train}^j$ 
\State $W_{test}^j \gets $ exposures from $W$ for the patients in the test set
\State $W_{train}^j, H_{train}^j \gets$ apply the chosen NMF method to $V_{train}^j$ with $K$ signatures
\State $c_j \gets cost(V_{test}^j,W_{test}^j H_{train}^j)$
\EndFor
\State \Return $c_{median} \gets median(c_1, \dots, c_J)$
\end{algorithmic}
\end{algorithm}

%%%%%%%%%%%%%%%%%%%%%%%%%%%%%%%%%%%%%%%%%%%
%%% Results
%%%%%%%%%%%%%%%%%%%%%%%%%%%%%%%%%%%%%%%%%%%

\section{Results} \label{sec:results_intro}
In this section we describe our results on both simulated and real data. For simulated data we present a study on Negative Binomial simulated data with different levels of dispersion where results from AIC, BIC, \SparseSignatures \citep{Lal2021}, \SigneR \citep{Rosales2017}, and \SignatureAnalyzer \citep{Tan2013} are compared with our proposed model selection procedure. These results are discussed in Section \ref{sec:results_simulations}, where we show that our method performs well and is robust to model misspecification. Our method is applied to the 21 breast cancer patients from \cite{alexandrov2013breastcancer} in Section \ref{sec:results_BRCA}, and to 286 prostate cancer patients from \cite{Campbell2020} in Section \ref{sec:results_Prostate}. The goodness of fit of the different models are evaluated using a residual analysis that shows a clear overdispersion with the Poisson model.

\subsection{Simulation study} \label{sec:results_simulations}
We simulate our data following the procedure of \cite{Lal2021} using the signatures from \cite{Tate2019}. We simulated 100 data sets for each scenario and varied the number of patients, the number of signatures and the model for the noise in the mutational count data. We considered $20,100$ and $300$ patients and either 5 or 10 signatures following \cite{RareSignatures2022} which states that the number of common signatures in each organ is usually between 5 and 10. For each simulation run we use signature 1 and 5 from \cite{Tate2019}, as they have been shown to be shared across all cancer types, and then we sample at random three or eight additional signatures from this set. The exposures are simulated from a Negative Binomial model with mean $6000$ and dispersion parameter $1.5$ as in \cite{Lal2021}. The mutational count data is then generated as the product of the exposures and signature matrix. Lastly, Poisson noise, Negative Binomial noise with dispersion parameter $\alpha \in \{10, 200\}$ or randomly sampled in $[10,500]$ are added to the mutational counts. The values of the patient specific dispersion are inspired from the data set in Section \ref{sec:results_BRCA}. A lower $\alpha$ is associated with higher dispersion, however the actual level of dispersion associated to a given $\alpha$ value depends on the absolute mutational counts as can be seen from the variance in Equation \eqref{eq:mean_var_nb}. Therefore it is not possible to directly compare these values with the ones estimated for the real data. 
\subsubsection{Simulation results}
\begin{figure}[H]
\centering
\subfigure[]{
    \resizebox{\textwidth}{!}{%
    \renewcommand{\arraystretch}{0}
\setlength{\fboxsep}{3mm} % box size
\setlength{\tabcolsep}{0pt}
\centering

\begin{tikzpicture}
\node (tab1){
     \begin{tabular}{ll *{5}{R}| *{5}{R}| *{5}{R}| *{5}{R} cc}
  \vspace{0.25cm} 
   & & \multicolumn{20}{c}{\textbf{True models}} & \\ \vspace{0.3cm}
   &  &  \multicolumn{5}{c}{Poisson} & \multicolumn{5}{c}{NB$_N (\alpha = 200)$} & \multicolumn{5}{c}{NB$_N (\alpha \text{ in } [10,500])$} & \multicolumn{5}{c}{NB$_N (\alpha = 10)$} \\
  \vspace{0.2cm}
  & \multicolumn{1}{l}{\#signatures} & \multicolumn{3}{c}{ \large 5} & \multicolumn{2}{c}{\large 10 } & \multicolumn{3}{c}{ \large 5} & \multicolumn{2}{c}{\large 10 }& \multicolumn{3}{c}{ \large 5} & \multicolumn{2}{c}{\large 10 }& \multicolumn{3}{c}{ \large 5} & \multicolumn{2}{c}{\large 10 } \\ \vspace{0.2cm}
  & \multicolumn{1}{l}{\#patients} & \multicolumn{1}{c}{\footnotesize 20} & \multicolumn{1}{c}{\footnotesize 100} & \multicolumn{1}{c}{\footnotesize300} & \multicolumn{1}{c}{ \footnotesize 100} & \multicolumn{1}{c}{ \footnotesize 300} & \multicolumn{1}{c}{\footnotesize 20} & \multicolumn{1}{c}{\footnotesize 100} & \multicolumn{1}{c}{\footnotesize300} & \multicolumn{1}{c}{ \footnotesize 100} & \multicolumn{1}{c}{ \footnotesize 300}& \multicolumn{1}{c}{\footnotesize 20} & \multicolumn{1}{c}{\footnotesize 100} & \multicolumn{1}{c}{\footnotesize300} & \multicolumn{1}{c}{ \footnotesize 100} & \multicolumn{1}{c}{ \footnotesize 300}& \multicolumn{1}{c}{\footnotesize 20} & \multicolumn{1}{c}{\footnotesize 100} & \multicolumn{1}{c}{\footnotesize300} & \multicolumn{1}{c}{ \footnotesize 100} & \multicolumn{1}{c}{ \footnotesize 300}  \\ \cline{3-22}
    & AIC & 1.00 &  1.00 & 1.00 & 0.00 & 1.00 & 0.00 & 0.00 & 0.00 & 0.00 & 0.00 & 0.00 & 0.00 & 0.00 & 0.00 & 0.00 & 0.00 & 0.00 & 0.00 & 0.00 & 0.00 & \multirow{3}{1cm}{\rotatebox{270}{Poisson}} & \multirow{6}{0.2cm}{\rotatebox{270}{\textbf{Assumed models}}}\\ 
  
   & BIC & 1.00 & 1.00 & 1.00 & 0.99 & 1.00 & 0.02 & 0.00 & 0.00 & 0.00 & 0.00 & 0.05 & 0.00 & 0.00 & 0.00 & 0.00 & 0.00 & 0.00 & 0.00 & 0.00 & 0.00 & \multirow{3}{1.5cm}{\rotatebox{270}{ }} \\ \vspace{0.2cm}
  
   & SigMoS & 0.92 & 1.00 & 1.00 & 0.86 & 0.87 & 0.93 & 0.99 & 1.00 & 0.88 & 0.85 & 0.68 & 0.86 & 0.95 & 0.64 & 0.70 & 0.65 & 0.32 & 1.00 & 0.44 & 0.45 \\
  \cline{3-22}
   & AIC  & 1.00 & 1.00 & 1.00 & 0.97 & 1.00 & 0.58 & 1.00 & 1.00 & 0.90 & 0.95 & 0.15 & 0.94 & 1.00 & 0.66 & 1.00 & 0.00 & 0.94 & 1.00 & 0.47 & 0.5 & \multirow{3}{1cm}{\rotatebox{270}{NB$_N$}} & \\ 
  
  & BIC & 1.00 & 1.00 & 1.00 & 1.00 & 1.00 & 1.00 & 1.00 & 1.00 & 0.99 & 1.00 & 0.76 & 1.00 & 1.00 & 0.88 & 1.00 & 0.00 & 0.93 & 0.95 & 0.11 & 0.2 \\ 
  
   & SigMoS \hspace{0.2cm} & 0.94 & 1.00 & 1.00 & 0.88 & 0.85 & 0.91 & 0.99 & 1.00 & 0.85 & 0.85 & 0.67 & 0.99 & 1.00 & 0.58 & 0.85 & 0.67 & 0.95 & 0.95 & 0.25 & 0.3  \\ \vspace{0.3cm} 
  
\end{tabular}
};     
\draw[draw=black] (-6.5,-3.5) rectangle ++(0.7,5);
\draw[draw=black] (-2.8,-3.5) rectangle ++(0.7,5);
\draw[draw=black] (0.94,-3.5) rectangle ++(0.7,5);
\draw[draw=black] (4.65,-3.5) rectangle ++(0.7,5);
\end{tikzpicture}}
    \label{fig:simulationresultsALL}
}
\subfigure[]{
    \centering
    \includegraphics[width = 0.9\textwidth]{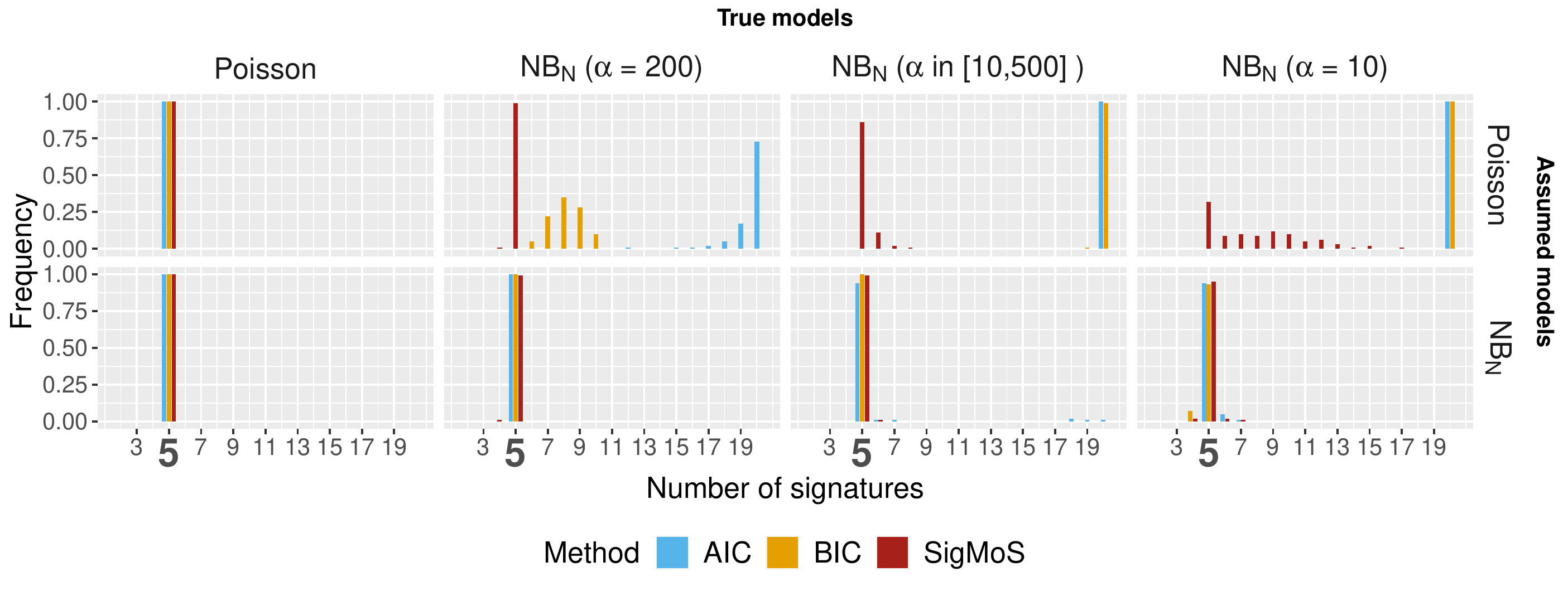} 
    \label{fig:simulationresults}
}

    \caption{Results from AIC, BIC, and SigMoS based on Po-NMF and NB$_N$-NMF using simulated data. Each method is applied on different simulated data sets for four different types of noise: Poisson and Negative Binomial with dispersion parameter $\alpha = 10, 200$ and $\alpha \sim U(10,500)$. \subref{fig:simulationresultsALL} The proportion of simulation runs where the number of signatures is correctly estimated. The true number of signatures varies in $\{5,10\}$ and the number of patients in $\{20,100,300\}$.  The rectangular boxes highlight the results shown in Figure \subref{fig:simulationresults}. The results are based on 100 simulation runs for scenarios with $20$ and $100$ patients and on 20 simulation runs for scenarios with $300$ patients.  \subref{fig:simulationresults} The estimated number of signatures in the range from 2 to 20 for 100 patients, where the true number of signatures is five. }
    \label{fig:simresboth}
\end{figure}

The effect of the model assumption on the estimated number of signatures using AIC, BIC (recall Equations \eqref{eq:AIC} and \eqref{eq:BIC}) and SigMoS as model selection procedures is shown in Figure \ref{fig:simresboth}. Figure \ref{fig:simulationresultsALL} summarizes results for all simulation studies and for each study, it displays the proportion of scenarios where the true number of signatures is correctly estimated from the different methods: the darker the green color the higher is this proportion. This shows that our proposed approach is estimating the number of signatures accurately and it is much more robust to model misspecifications compared to AIC and BIC. For example, when the true model has a small dispersion of $\alpha = 200$ and the Poisson model is assumed, the difference between the performance of SigMoS and of AIC and BIC is already substantial. Here, AIC and BIC are never estimating the true number of signatures correctly, whereas our SigMoS procedure estimates the correct number of signatures in most cases ($\geq 85\%$). The table also shows that the higher the dispersion in the model, the harder it is to estimate the true number of signatures even when the correct model is specified.

Figure \ref{fig:simulationresults} depicts the actual estimated number of signatures in the range from 2 to 20 for the 100 data sets with 5 signatures and 100 patients. This clearly shows that the higher the overdispersion in the model, the more is the number of signatures overestimated. Assuming Poisson in the case of $\alpha = 200$ we see that AIC is already overestimating the number of signatures. Here, these additional signatures are needed to explain the noise that is not accounted for by the Poisson model.  Having an even higher overdispersion makes both AIC and BIC highly overestimate the number of signatures to a value that is plausibly much higher than 20. Even high overdispersion does not influence our SigMoS procedure in the same way and our approach is still estimating the true number of signatures for a large proportion of the scenarios. Assuming the Negative Binomial model all of the three methods have a really high performance, as the Negative Binomial accounts for both low and high dispersion.

In the simulation study from Figure \ref{fig:simulationresults} we also consider the accuracy of the MLE for the $\alpha$ value in the two scenarios where each patient has the same $\alpha$. Our approach estimates the true $\alpha$ with high accuracy when the dispersion is high i.e.\, $\hat{\alpha} \in [9.21, 11.78]$ for $\alpha = 10$, $\alpha$ is slightly overestimated when the dispersion is low: for $\alpha = 200$ we find $\hat{\alpha} \in [225.8, 292.7]$.
However according to Figure \ref{fig:simulationresults} this does not affect the performance of our model selection procedure.  

\subsubsection{Method comparison}
Several methods have been proposed in the literature for estimating the number of signatures in cancer data. In the following we present the results of a comparison between our method and three commonly used methods in the literature: \SparseSignatures, \SignatureAnalyzer, and \SigneR. \SparseSignatures \citep{Lal2021} provides an alternative cross-validation approach where the test set is defined by setting $1\%$ of the entries in the count matrix to $0$. Then NMF is iteratively applied to the modified count matrix and the entries are updated at each iteration. The resulting signature and exposure matrices are used to predict the entries of the matrix corresponding to the test set. \SignatureAnalyzer \citep{Tan2013}, on the other hand, proposes a procedure where a Bayesian model is used and maximum a posteriori estimates are found with a majorize-minimization algorithm. Lastly, with \SigneR \citep{Rosales2017} an empirical Bayesian approach based on BIC is used to estimate the number of mutational signatures.

For our method comparison, we run all methods on the simulated data from Figure \ref{fig:simulationresults}. For each method and simulation setup we only allow the number of signatures to vary from two to eight due to the long running time of some of these methods. 

\begin{figure}[h]
    \centering
    \includegraphics[width = \textwidth]{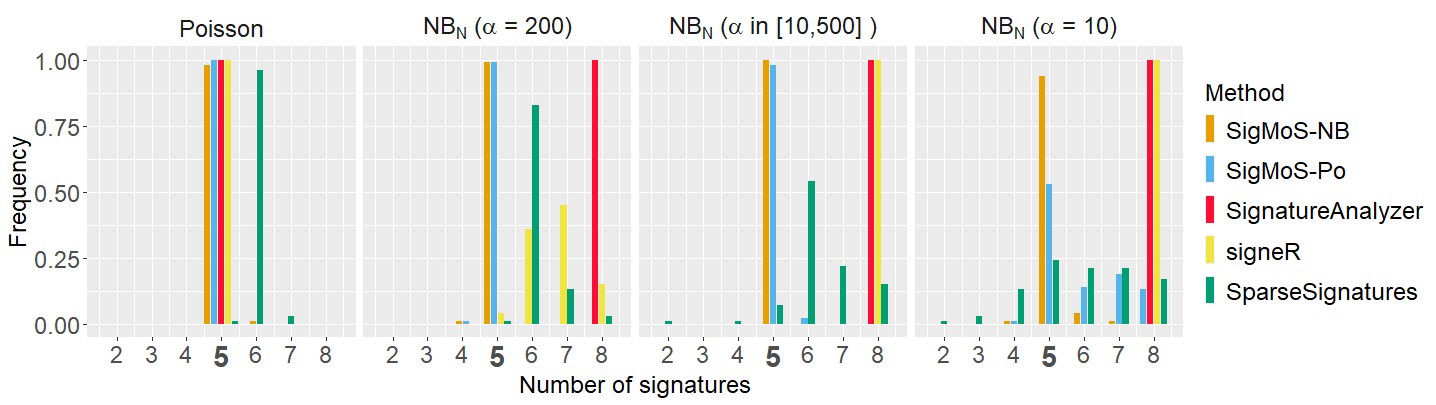}
    \vspace{-2em}
    \caption{Method comparison using simulated data. Each method is applied on the data sets from Figure \ref{fig:simulationresults} and, for each data set, the value of the estimated number of signatures is kept. We test values for the number of signatures from two to eight for Poisson noise and Negative Binomial noise with $\alpha = 10, 200$, and a patient specific dispersion parameter $\alpha \sim U(10,500)$.}
    \label{fig:methodcomparison_simulations}
\end{figure}

Figure \ref{fig:methodcomparison_simulations} shows that, when Poisson data are simulated all methods have a very good performance and can recover the true number of signatures in most of the simulations. The poor performance of \SparseSignatures could be affected by not having a fixed background signature. Indeed, the improved performance of \SparseSignatures when a background signature is included has also been shown in \cite{Lal2021}. When Negative Binomial noise is added to the simulated data with a moderate dispersion ($\alpha = 200$), however, both \SignatureAnalyzer and \SigneR have low power emphasizing the importance of correctly specifying the distribution for these methods, whereas our proposed approach (regardless of the distributional assumption) and \SparseSignatures maintain good power. For patient specific dispersion also the power of \SparseSignatures decreases. Good performance is also achieved with our proposed approach under high dispersion ($\alpha = 10$) if the correct distribution is assumed. These results demonstrate that SigMoS is accurate for detecting the correct number of signatures and it performs well also in situations with overdispersion compared to other methods.

\subsection{Breast Cancer Data} \label{sec:results_BRCA}
This data set consists of the mutational counts from the 21 breast cancer patients that has previously been described and analyzed in several papers \citep{Nik-Zainal2012, alexandrov2013breastcancer, Fischer2013}. The data can be found through the link \url{ftp://ftp.sanger.ac.uk/pub/cancer/AlexandrovEtAl} from \cite{alexandrov2013signatures}.

\begin{figure}[h]
\centering
\subfigure[Estimated number of signatures]{
\resizebox{0.6\textwidth}{!}{%
\begin{tabular}{ccc}
\toprule
                         & \multicolumn{2}{c}{Assumed models}                      \\ \cline{2-3}
Model selection procedure \hspace{0.5em}  & \multicolumn{1}{l}{Po-NMF} & \multicolumn{1}{l}{NB-NMF} \\ \hline
\multicolumn{1}{c}{SigMoS}                    & 3                         & 3                          \\
\multicolumn{1}{c}{BIC}                       & 6                         & 3                          \\ \bottomrule
       \vspace{0.05cm}
\end{tabular}
    }
    \label{fig:BRCAtable}
}

\subfigure[Model fit]{
\includegraphics[width = 0.7\textwidth]{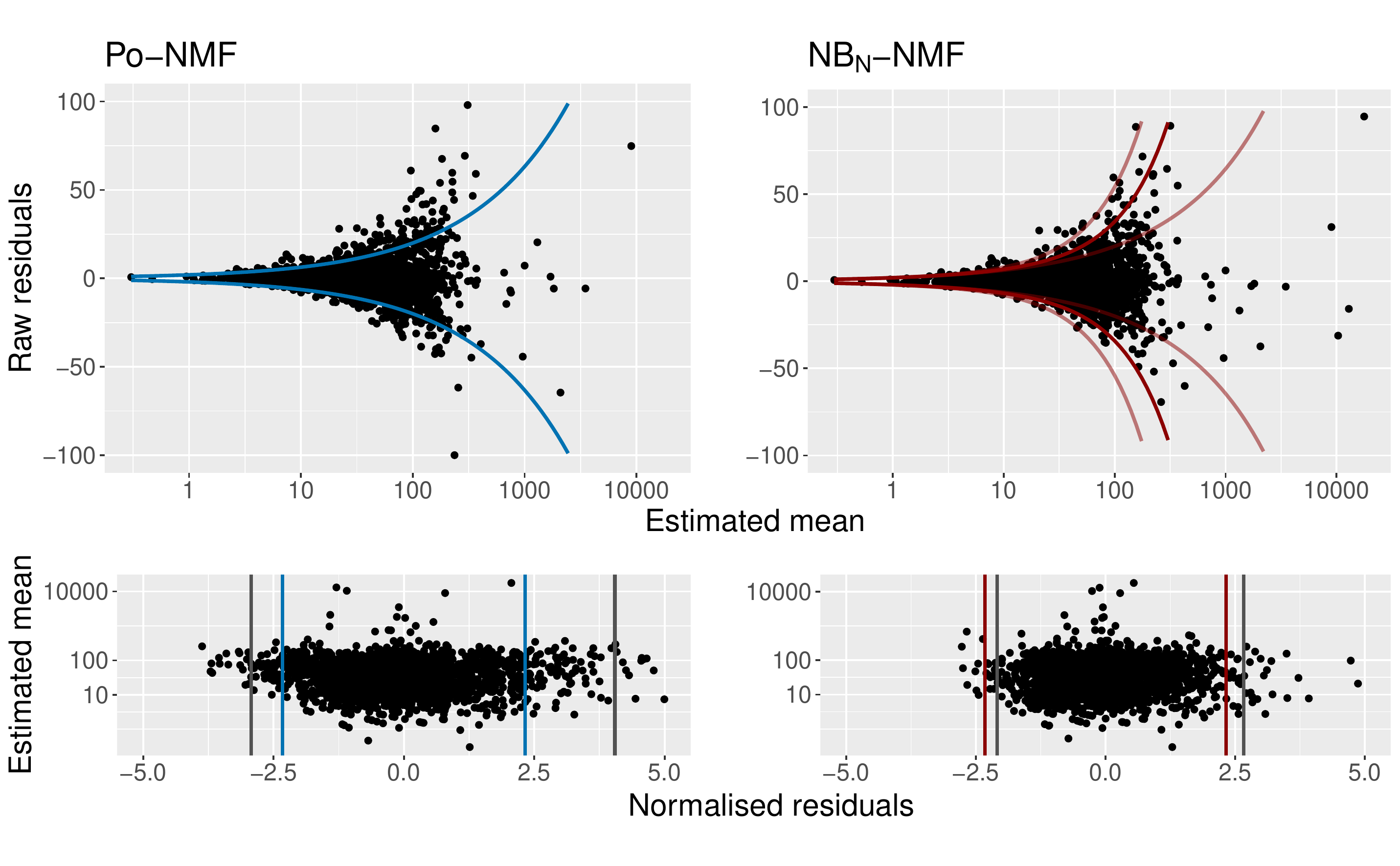}
 \label{fig:BRCAresidual}
}

\caption{Results for Po-NMF and NB$_\text{N}$-NMF applied to a data set with 21 breast cancer patients. \subref{fig:BRCAtable} The optimal number of signatures estimated from SigMoS and BIC when using Po-NMF and NB$_\text{N}$-NMF. \subref{fig:BRCAresidual} The residual plots for Po-NMF and NB$_\text{N}$-NMF when assuming the estimated number of signatures from SigMoS i.e. 3 signatures in both cases. The lines in the top plot correspond to two times the expected variance under the chosen distributional assumption. As the NB$_\text{N}$-NMF holds 21 different expected variances, we have chosen to plot the median, minimum and maximum variance among the 21. The second plots show the normalized residuals. The vertical blue and red lines depict the theoretical quantiles and the gray lines show the observed quantiles.}
\label{fig:BRCA21}
\end{figure}

In Figure \ref{fig:BRCAtable}, we have applied SigMoS and BIC to choose the number of signatures for both Po-NMF and NB$_\text{N}$-NMF. We have included the BIC to compare with the SigMoS method as it provides similar results to the state-of-the-art methods. SigMoS indicates to use three signatures for both methods. This is in line with the results of our simulation study, where we show that our model selection is robust to model misspecification. According to BIC, six signatures are needed for Po-NMF whereas only three signatures should be used with NB$_\text{N}$-NMF which emphasizes the importance of a correct model choice when using BIC.

For three signatures we show in Figure \ref{fig:BRCAresidual} the corresponding raw residuals $R_{nm} = V_{nm} - (WH)_{nm}$ to determine the best fitting model. The residuals are plotted against the expected mean $(WH)_{nm}$, as the variance in both the Poisson and Negative Binomial model depends on this value. The colored lines in the residual plots correspond to $\pm 2\sigma$ for the Poisson and the Negative Binomial distribution respectively. The variance $\sigma^2$ can be derived from Equation \pef{eq:mean_var_nb} for the Negative Binomial model and is equal to the mean for the Poisson model. 

For Po-NMF we observe a clear overdispersion in the residuals, which suggests to use a Negative Binomial model. In the residual plot for the NB$_\text{N}$-NMF we see that the residuals have a much better fit to the variance structure, which is indicated by the colored lines. The quantile lines in the lower panel with normalized residuals again show that the quantiles from the NB$_\text{N}$-NMF are much closer to the theoretical ones, suggesting that the Negative Binomial model is better suited for this data. The patient specific dispersion is very diverse in this data as the last patient has $\alpha_{21} = 26083$ and the $\alpha$ values for the rest of the patients are between 16 and 550.

\subsection{Prostate Cancer Data} \label{sec:results_Prostate}
\begin{figure}[h]
\centering
\subfigure[Estimated number of signatures]{
\resizebox{0.6\textwidth}{!}{%
\begin{tabular}{ccc}
\toprule
                         & \multicolumn{2}{c}{Assumed models}                      \\ \cline{2-3}
Model selection procedure \hspace{0.5em}  & \multicolumn{1}{l}{Po-NMF} & \multicolumn{1}{l}{NB-NMF} \\ \hline
\multicolumn{1}{c}{SigMoS}  & 6    & 5     \\
\multicolumn{1}{c}{BIC}   & 9      & 4     \\ \hline
       \vspace{0.05cm}
\end{tabular}
    }
    \label{fig:Prostatetable}
}

\subfigure[Model fit]{
\includegraphics[width = 0.7\textwidth]{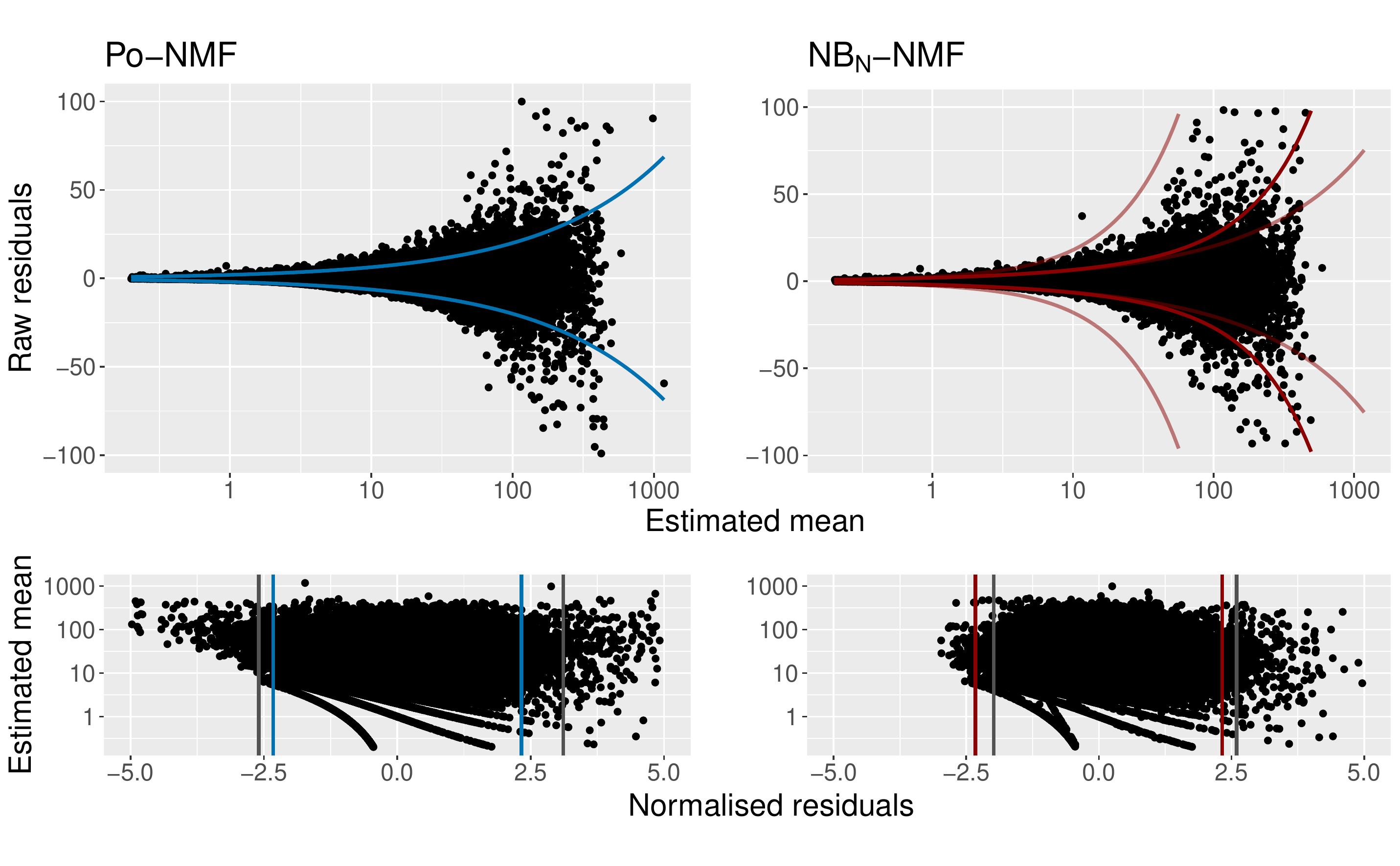}
 \label{fig:Prostateresidual}
}

\caption{Results for Po-NMF and NB$_\text{N}$-NMF applied to a data set with 286 prostate cancer patients from the PCAWG database \citep{Campbell2020}. \subref{fig:Prostatetable} The optimal number of signatures estimated from SigMoS and BIC when using Po-NMF and NB$_\text{N}$-NMF. \subref{fig:Prostateresidual} The residual plots for Po-NMF and NB$_\text{N}$-NMF when assuming the estimated number of signatures from SigMoS i.e. 5 and 6 signatures. The lines in the first plot correspond to two times the expected variance under the chosen distributional assumption. For NB$_\text{N}$-NMF, the colored lines in the top plot show the median, minimum and maximum variance among the patients. The bottom plots show the normalized residuals. The vertical blue and red lines depict the theoretical quantiles and the gray lines the observed quantiles.}
\label{fig:PCAWG_prostate}
\end{figure}

We also considered a more recent data set from the Pan-Cancer Analysis of Whole Genomes (PCAWG) database \citep{Campbell2020} where 2782 patients from different cancer types are available. The mutational counts from the full PCAWG database can be found at \url{https://www.synapse.org/#!Synapse:syn11726620}. From this data set, we extracted mutational counts for all the 286 prostate cancer patients and used them directly for our analysis. We chose again both the Poisson and Negative Binomial as underlying distributions for the NMF and in both cases we applied SigMoS for determining the number of signatures. We present the results in Figure \ref{fig:PCAWG_prostate}. Figure \ref{fig:Prostatetable} shows again that our model selection procedure is more stable under model misspecification compared to BIC: the estimated number of signatures is changing from 9 to 4 between the two model assumptions for BIC, but only from 6 to 5 for SigMoS. As for Figure \ref{fig:BRCAresidual}, the residuals in Figure \ref{fig:Prostateresidual} show that the NB$_\text{N}$-NMF model provides a much better fit to the data than the Po-NMF. The estimated values for the patient specific dispersion are $\alpha_n \in [1.4,4279]$ with a median of $140$. 

%%%%%%%%%%%%%%%%%%%%%%%%%%%%%%%%%%%%%%%%%%%
%%% Discussion
%%%%%%%%%%%%%%%%%%%%%%%%%%%%%%%%%%%%%%%%%%%

\section{Discussion} \label{sec:discussion}
Mutational profiles from cancer patients are a widely used source of information and NMF is often applied to these data in order to identify signatures associated with cancer types. We propose a new approach to perform the analysis and signature extraction from mutational count data where we emphasize the importance of validating the model using residual analysis, and we propose a robust model selection procedure.

We use the Negative Binomial model as an alternative to the commonly used Poisson model as the Negative Binomial can account for the high dispersion in the data. As a further extension of this model, we allow the Negative Binomial to have a patient specific variability component to account for heterogeneous variance across patients.

We propose a model selection approach for choosing the number of signatures. As we show in Section \ref{sec:results_simulations} this method works well with both Negative Binomial and Poisson data and it is a robust procedure for choosing the number of signatures. We note that the choice of the divergence measure for the $cost$ function in Algorithm \ref{alg:crossval} is not trivial and may favor one or the other model and thus a comparison of the costs between different NMF methods is not possible. For example, in our framework, we use the Kullback-Leibler divergence which would favor the Poisson model. This means that a direct comparison between the cost values for Po-NMF and NB$_\text{N}$-NMF is not feasible. To check the goodness of fit and choose between the Poisson model and the Negative Binomial model we propose to use the residuals instead. 

We investigated the role of the cost function in our model selection by including the Frobenius norm and the Beta and Itakura-Saito (IS) \citep{Fevotte2011} divergence measures from \cite{Li2012} where the authors propose a fast implementation of the NMF algorithm with general Bregman divergence. In this investigation the cost function did not influence the optimal number of signatures. The only difference was how the cost values differed among the NMF methods, as each cost function favored the models differently. Therefore we chose to use the Kullback-Leibler divergence and compared the methods with the residual analysis. 

Less signatures are found when accounting for overdispersion with the Negative Binomial model. Indeed, there is no need to have additional signatures explaining noise, which we assume is the case for the Poisson model. We show that the Negative Binomial model is more suitable and therefore believe the corresponding signatures are more accurate. This can be helpful when working with mutational profiles for being able to better associate signatures with cancer types and for a clearer interpretation of the signatures when analyzing mutational count data. For example, the recent results in \cite{RareSignatures2022} use a large data set with several different cancer types and show that there exists a set of common signatures that is shared across organs and a set of rare signatures that are only found with a sufficiently large sample size. To recover the common signatures the patients with unusual mutational profiles were excluded as they are introducing additional variance in the signature estimation procedure. We speculate that changing the Poisson assumption in this approach with the Negative Binomial distribution could provide a simpler and more robust way to extract common signatures. Indeed, the Negative Binomial model allows for more variability in the data and our simulation results and residual plots in Section \ref{sec:results_intro} show that the Negative Binomial distribution is beneficial for stable signature estimation.

The workflow for analyzing the data, and the procedures in Algorithms \ref{alg:nbnmf_alphan} and \ref{alg:crossval} are available in the R package SigMoS at \url{https://github.com/MartaPelizzola/SigMoS}.

%%%%%%%%%%%%%%%%%%%%%%%%%%%%%%%%%%%%%%%%%%%
%%% Methods
%%%%%%%%%%%%%%%%%%%%%%%%%%%%%%%%%%%%%%%%%%%

\section{Methods}\label{sec:methods}
In Section \ref{sec:NegBinNMF} we describe the negative binomial NMF model applied to mutational count data and we propose an extension where a patient specific dispersion coefficient is used. The majorization-minimization (MM) procedure for patient specific dispersion $\{ \alpha_1, \dots, \alpha_N \}$ can be found in Section \ref{subsec:patientNBNMF}. In our application, we propose to use Negative Binomial maximum likelihood estimation for $\alpha$ (see Section \ref{subsec:negbin}) and $\{ \alpha_n: 1 \leq n \leq N \}$ (see Section \ref{subsec:patientNBNMF}) instead of the grid search approach adopted in \cite{Gouvert2020}. The pseudocode shown in the initial steps of Algorithm \ref{alg:nbnmf_alphan} describes this approach for patient specific dispersion. For shared dispersion among all patients and mutation types we simply set $\alpha = \alpha_1 = \cdots = \alpha_N$ in Algorithm \ref{alg:nbnmf_alphan}.

\subsection{Patient specific NB$_\text{N}$-NMF}\label{sec:methods_alphai}
As we discuss in Section \ref{sec:results_BRCA} the variability in mutational counts among different patients can be really high. Thus we extend the Negative Binomial NMF from \cite{Gouvert2020} (see Section \ref{subsec:negbin}) by including a patient specific component (see Section \ref{subsec:patientNBNMF}). We noticed that the variability among different patients is usually much higher than the one among different mutation types, thus we decided to focus on patient specific dispersion.

The entries in $V$ are modeled as 
\[V_{nm} \sim \text{NB}\left( \alpha_n, \frac{(WH^T)_{nm}}{\alpha_n + (WH^T)_{nm}}\right), \]
where $\alpha_n$ is the dispersion coefficient of each patient, and the corresponding Gamma-Poisson hierarchical model can be rewritten as:
\begin{align} \label{eq:gamma_patient}
    V_{nm}|a_{nm} \sim \text{Po}(a_{nm}(WH)_{nm}) \\ \nonumber
    a_{nm} \sim \text{Gamma}(\alpha_n, \alpha_n).
\end{align}
Here $a_{nm}$ is the parameter responsible for the variability in the Negative Binomial model. Note that $ \mathbb{E}[a_{nm}] = 1$ and $Var(a_{nm}) = 1/\alpha_n$.

Now we can write the Negative Binomial log-likelihood function with patent specific $\alpha_n$
\begin{align} \label{eq:fulllik}
    \ell(W,H;V) & = \sum_{n=1}^N \sum_{m=1}^M \Bigg\{ \log { \binom{\alpha_n + V_{nm}  - 1}{\alpha_n}} + V_{nm} \log \left( { \frac{(WH)_{nm}}{\alpha_n + (WH)_{nm}}}\right)  \\ \nonumber
    &  + \alpha_n \log \left( { 1 - \frac{(WH)_{nm}}{\alpha_n + (WH)_{nm}} } \right) \Bigg\}
\end{align} 
and recognize the negative of the log-likelihood function as proportional to the following divergence:
\begin{align}
    d_N(V||WH) & = \sum_{n=1}^N \left \{  \sum_{m=1}^M V_{nm} \log \left(\frac{V_{nm}}{ (WH)_{nm}}\right) - (\alpha_n + V_{nm}) \log \left(\frac{\alpha_n + V_{nm}}{\alpha_n + (WH)_{nm}} \right) \right \} \label{eq:div_methods}
\end{align} 
assuming fixed $\alpha_1, \cdots, \alpha_N$.
The term $\log { \binom{\alpha_n + V_{nm}  - 1}{\alpha_n}}$ in the likelihood is a constant we can remove and then we have added the constants $V_{nm} \log (V_{nm})$, $\alpha_{n} \log (\alpha_{n})$ and $(V_{nm} + \alpha_n)  \log (V_{nm} + \alpha_n)$.  

%%%%%%%%%%%%%%%%%%%%
Following the steps in \cite{Gouvert2020}, we will update $W$ and $H$ one at a time, while the other is assumed fixed. We will show the procedure for updating $H$ using a fixed $W$ and its previous value $H^t$. First we construct a majorizing function $G(H, H^t)$ for $d_N(V||WH)$ with the constraint that $G(H, H) = d_N(V||WH)$.  The first term in Equation \eqref{eq:div_methods} can be majorized using Jensen's inequality leading to
\begin{align}
    d_N(V||WH) & = \sum_{n=1}^N \sum_{m=1}^M \left \{ V_{nm} \log \left(\frac{V_{nm}}{\sum_{k=1}^K W_{nk}H_{km}} \right) - (\alpha_n + V_{nm}) \log \left(\frac{\alpha_n + V_{nm}}{\alpha_n + \sum_{k=1}^K W_{nk}H_{km}} \right) \right \} \\ \nonumber 
    & \leq \sum_{n=1}^N \sum_{m=1}^M\Bigg\{ V_{nm}  \log V_{nm} - V_{nm} \sum_{k=1}^K \beta_{k} \log \frac{W_{nk}H_{km}}{\beta_{k}} \\ \nonumber 
    &  + (\alpha_n + V_{nm}) \log \left(\frac{\alpha_n + \sum_{k=1}^K W_{nk}H_{km}}{\alpha_n + V_{nm}} \right) \Bigg\} \label{eq:jensen}
\end{align} 
where $\beta_{k} = \nicefrac{W_{nk}H_{km}^t}{\sum_{k=1}^K W_{nk}H^t_{km}}$. The second term can be majorized with the tangent line using the concavity property of the logarithm:
\begin{align}
    d_N(V||WH) & = \sum_{n=1}^N \sum_{m=1}^M \Bigg\{ V_{nm} \log V_{nm} - V_{nm} \sum_{k=1}^K \beta_{k} \log \frac{W_{nk}H_{km}}{\beta_{k}}  \\ \nonumber
    & + (\alpha_n + V_{nm}) \log \left(\frac{\alpha_n + \sum_{k=1}^K W_{nk}H_{km}}{\alpha_n + V_{nm}} \right) \Bigg\} \\ \nonumber
    & \leq \sum_{n=1}^N \sum_{m=1}^M \Bigg\{ V_{nm} \log V_{nm} - V_{nm} \sum_{k=1}^K \beta_{k} \log \frac{W_{nk}H_{km}}{\beta_{k}} \\ \nonumber 
    & + (\alpha_n + V_{nm}) \log \left(\frac{\alpha_n + (WH^t)_{nm}}{\alpha_n + V_{nm}} \right) + \frac{W_{nm}}{\alpha_n + (WH^t)_{nm}}(H_{nm} - H^t_{nm}) \Bigg\} = G(H, H^t).
    \label{eq:tangent}
\end{align} 
Lastly, we need to show that $G(H, H) = d_N(V||WH)$. This follows from
\begin{align}
   G(H, H) = &  \sum_{n=1}^N \sum_{m=1}^M  \Bigg\{ V_{nm} \log V_{nm}  - V_{nm} \sum_{k=1}^K \beta_{k} \log \frac{W_{nk}H_{km}}{\beta_{k}} \\ \nonumber 
    &  + (\alpha_n + V_{nm}) \log \left(\frac{\alpha_n + (WH)_{nm}}{\alpha_n + V_{nm}} \right) + \frac{W_{nm}}{\alpha_n + (WH)_{nm}}(H_{nm} - H_{nm}) \Bigg\} \\ \nonumber
    & = \sum_{n=1}^n \sum_{m=1}^M \Bigg\{ V_{nm} \log V_{nm}  - V_{nm}\sum_{k=1}^K \frac{W_{nk}H_{km}}{\sum_{k=1}^K W_{nk}H_{km}} \log \frac{W_{nk}H_{km}}{\frac{W_{nk}H_{km}}{\sum_{k=1}^K W_{nk}H_{km}}} \\ \nonumber 
    &  - (\alpha_n + V_{nm}) \log \left(\frac{\alpha_n + V_{nm}}{\alpha_n + \sum_{k=1}^K W_{nk}H_{km}} \right) \Bigg\} \\ \nonumber 
    & =  \sum_{n=1}^n \sum_{m=1}^M \Bigg\{ V_{nm} \log V_{nm} - V_{nm} \cdot 1 \cdot \log \left(\sum_{k=1}^K W_{nk}H_{km} \right) \\ \nonumber 
    &- (\alpha_n + V_{nm}) \log \left(\frac{\alpha_n + V_{nm}}{\alpha_n + \sum_{k=1}^K W_{nk}H_{km}} \right) \Bigg\} \\ \nonumber 
    &  = \sum_{n=1}^N  \sum_{m=1}^M \Bigg\{ V_{nm} \log \left(\frac{V_{nm}}{\sum_{k=1}^K W_{nk}H_{km}}\right) - (\alpha_n + V_{nm}) \log \left(\frac{\alpha_n + V_{nm}}{\alpha_n + \sum_{k=1}^K W_{nk}H_{km}} \right) \Bigg\} \\ \nonumber 
    &  =  d_N(V||WH)
\end{align} 

Having defined the majorizing function $G(H, H^t)$ in \eqref{eq:tangent}, we can derive the following multiplicative update for $H$:
\begin{align}
   H_{km}^{t+1} = H_{km}^t \frac{\sum_{n=1}^N \frac{V_{nm}}{(WH^t)_{nm}} W_{nk}}{\sum_{n=1}^N \frac{V_{nm} + \alpha_n}{(WH^t)_{nm} + \alpha_n} W_{nk}}. \label{eq:updateH_methods}
\end{align}
Similar calculations can be carried out for $W$ to obtain the following update:
\begin{align}
    W^{t+1}_{nk} = W^t_{nk} \frac{\sum_{m=1}^M \frac{V_{nm}}{(W^tH)_{nm}} H_{km}}{\sum_{m=1}^M \frac{V_{nm} + \alpha_n}{(W^tH)_{nm} + \alpha_n} H_{km}}.  \label{eq:updateW_methods}
\end{align}
It is straightforward to see that when $\alpha_n = \alpha$ for all $ n = 1, \dots, N$ then the updates for $W$ and $H$ equal those in \cite{Gouvert2020}. Additionally, as shown in \cite{Gouvert2020} when $\alpha \to \infty$ the updates of the Po-NMF \citep{Lee1999} are recovered. 
The pseudo code in Algorithm \ref{alg:nbnmf_alphan} summarizes the NB$_\text{N}$-NMF model discussed in this section. 

\subsection{Code for method comparison}
For \SparseSignatures we use the function \texttt{nmfLassoCV} with \texttt{normalize\_counts} being set to FALSE and \texttt{lambda\_values\_alpha} and \texttt{lambda\_values\_beta} to zero. All the other parameters are set to their default values. When applying \SignatureAnalyzer we used the following command \texttt{python SignatureAnalyzer-GPU.py --data f --prior\_on\_W L1 --prior\_on\_H L2 --output\_dir d --max\_iter 1000000 --tolerance 1e-7 --K0 8}. For \SigneR we used the default options.

\section*{Acknowledgement}
We would like to thank Simon Opstrup Drue for helpful comments and suggestions on an earlier version of this manuscript. The research reported in this publication is supported by the Novo Nordisk Foundation. MP acknowledges funding of the Austrian Science Fund (FWF Doctoral Program Vienna Graduate School of Population Genetics", DK W1225-B20).

\bibliographystyle{apalike}
\bibliography{bibliography}
\end{document}